\documentclass[aps,pre,twocolumn,nofootinbib]{revtex4-2}

\usepackage[tbtags]{amsmath}
\usepackage{amssymb}
\usepackage{mathrsfs}
\usepackage{hyperref}
\hypersetup{
 colorlinks=true,
 linkcolor=blue,
 citecolor=blue,
 urlcolor=blue
}
\usepackage{graphicx}
\usepackage{bm}

\newcommand{\br}{{\bm{r}}}
\newcommand{\bp}{{\bm{p}}}
\newcommand{\bs}{{\bm{s}}}
\newcommand{\bl}{{\bm{l}}}
\newcommand{\bj}{{\bm{j}}}
\newcommand{\bM}{\mathsf{M}}
\newcommand{\sH}{{\mathscr{H}}}

\newcommand{\kbar}{\mathring{\kappa}}
\newcommand{\sord}{\mathring{s}}
\newcommand{\lord}{\mathring{l}}
\newcommand{\jord}{\mathring{\jmath}}
\def\<{\langle}
\def\>{\rangle}
\renewcommand{\Re}{\mathop{\rm Re}}

\newcommand{\eps}{\varepsilon}
\newcommand{\NN}{\nonumber\\}
\newcommand{\pp}[2]{\frac{\partial #1}{\partial #2}}
\newcounter{myfig}
\newcommand{\putfig}[4]{\begin{figure}[#1]
 \centering
 \stepcounter{myfig}
 \includegraphics[width=#2\linewidth]{Fig\arabic{myfig}.eps}
 \caption{\label{fig:#3}#4}
 \end{figure}}
\newcommand{\mtrx}[1]{\begin{pmatrix}#1\end{pmatrix}}

\begin{document}
\title{Classical periodic orbits in extended phase space
for spherical harmonic oscillator with spin-orbit coupling}
\author{Ken-ichiro Arita}
\email{arita@nitech.ac.jp}
\affiliation{Department of Physical Science and Engineering,
Nagoya Institute of Technology, Nagoya 466-8555, Japan}
\date{\today}

\begin{abstract}
A complete analysis of classical periodic orbits (POs) and
their bifurcations was conducted in spherical harmonic
oscillator system with spin-orbit coupling.  The motion of the spin is
explicitly considered using the spin canonical variables derived by
semiclassical approximation to the spin coherent state path-integral
representation.  In addition to the diametric and two circular PO
families with frozen spin, solutions that bridge two circular POs are
found in which orbital motion is coupled to spin precession.
In addition, each bridge encounters a secondary bifurcation on the way
from one circular PO to the other and generates a new PO that survives
at higher energies while maintaining a constant period.  The generic
expressions for those POs are obtained explicitly, and all the above
peculiar bifurcation scenarios are described fully analytically.
\end{abstract}
\maketitle

\section{Introduction}
\label{sec:introduction}

Harmonic oscillator (HO) system is one of the fundamental systems both
in quantum and classical dynamics.  Three-dimensional isotropic HO
system shows peculiar simplicity due to the dynamical SU(3) symmetry.
In quantum mechanics, it has strongly degenerate equidistant energy
levels, and in classical mechanics, all the trajectories are
periodic.  This system is still integrable and exactly solvable after
inclusion of spin-orbit coupling given by a simple scalar product of
spin and orbital angular momentum vectors.  Such Hamiltonian is
expressed as
\begin{equation}
H=\frac{\bp^2}{2M}+\frac{M\omega^2\br^2}{2}-\kappa \bs\cdot(\br\times\bp).
\label{eq:Hamiltonian}
\end{equation}
Here $M$ is the particle mass, $\omega$ is the oscillator frequency, and
$\kappa$ is the spin-orbit coupling strength.  This Hamiltonian
appears as a simple model for the nuclear mean
field\cite{Nil55,RSText}.  This type of spin-orbit coupling appears in
the nonrelativistic approximation to the Dirac equation for spin
$1/2$ charged fermions under Coulomb field.  In the nuclear mean
field, strong spin-orbit coupling is known to exist for both protons
and neutrons which comes from a different origin than the case of the Dirac
equation.  It plays a significant role in explaining the nuclear ``magic
numbers'' in more realistic mean field
potentials\cite{MJ55,BMText1,RSText}.

In analyzing single-particle shell structures, namely, gross
structures of fluctuations in the single-particle level densities,
semiclassical periodic-orbit theory (POT)
\cite{Gut71,BaBlo72,GutText,TabText,OzoText,BBText}
is very useful.  It represents the fluctuating part of the level
density as the contribution of classical periodic orbits (POs) in the
corresponding classical Hamiltonian system.  This POT was applied
to analyze the gross shell structures in nuclear and cluster
systems\cite{StrMag76,StrMag77,NisMot90,Fri90,Ari16}.

If one wants to apply POT to the system with spin degree of
freedom, one has to define the classical counterpart of the spin,
which is purely quantum mechanical object.  To treat
spin-orbit coupling in semiclassical POT, the
Wentzel-Kramers-Brillouin (WKB)
method was first proposed by Littlejohn and
Flynn\cite{LitFly92} in which the multicomponent wave equation was
diagonalized using the Wigner-Weyl calculus and was reduced to a set of
uncoupled equations corresponding to the individual spin
channels\cite{YabHor87,LitFly91}.  This is a kind of adiabatic
approximation which treats spin as slow variable.
However, it suffers a serious problem of mode conversions where
adiabatic approximation breaks down and the classical motion
becomes singular.  A diabatic correction at the mode conversion point
has been proposed\cite{FriGuh93} which seems to be appropriate in
some numerical applications, however, it remains ambiguous and lacks
theoretical basis\cite{AmaBra02}.

In another prescription, the spin degree of freedom is explicitly
incorporated into the equations of motion (EOM).  The classical phase
space is extended by introducing a spin canonical variables.  The
classical dynamics of the spin are derived by a semiclassical
approximation of the spin coherent state path integral
representation\cite{KurSuz80}.  Along with this prescription,
the semiclassical trace formula for system with spin degree of freedom
was derived\cite{PleAma02,PleZai03} in which spin dynamics was
explicitly considered in the classical motion.
Although the semiclassical
approximation was justified only for large spins, 
the trace formula was found to be applicable to the system of particles
with small spins, such as electrons and nucleons, at least
for the Hamiltonians which depend linearly on the spin.
The Hamiltonian (\ref{eq:Hamiltonian}) certainly meets this requirement.
This finding increases the significance of investigating classical motion,
especially the POs, in the extended phase space
where spin motion is taken explicitly in the EOM.

Apart from its physical importance as the nuclear mean-field model,
the Hamiltonian (\ref{eq:Hamiltonian}) provides us an interesting and
significant example of the integrable spin-orbit-coupled system.
The classical motions in the same Hamiltonian
were investigated by Rozmej and Arview\cite{RozArv92}.
The main purpose of their work was to specify the origin of chaotic
nature in the nuclear mean-field model.  They found that the classical
motion in the anisotropic HO system, which is integrable,
turns gradually chaotic with increasing spin-orbit coupling strength,
and concluded that the spin-orbit coupling can be regarded as one
possible source of nuclear chaoticity.
In their paper, they also studied the classical motions in the
spherical potential, only numerically by the Poincar\'{e} plot for
two-dimensional orbits in the angular-momentum projected space.
The result indicates the integrability of the classical system,
but the analytical studies of the orbit have not been attempted.

Classical-quantum correspondence in the Hamiltonian similar to
Eq.~(\ref{eq:Hamiltonian}) was considered by Amann et al. in
Ref.~\cite{AmaBra02}.  They incorporated anisotropy into the HO potential
and applied the WKB method.  A good quantum-classical correspondence
was found in such an anisotropic case.  However, in the spherical case,
some POs lie in the mode conversion point and the WKB method cannot
be applied.  They also found several
inconsistencies between contribution of classical POs and the exact
quantum spectrum.  Then they got interest in applying the
trace formula in the extended phase space they develped, but actual
application has not yet been made.  Some details on this
will be discussed later in Sec.~\ref{sec:summary}.

In this paper, I give the general analytic solution to the EOM for
Hamiltonian (\ref{eq:Hamiltonian}) by explicitly taking account of the
spin degree of freedom using the classical spin variables.  The EOM
are nonlinear in orbital and spin variables, but the spin motion can
be exactly solved in the spherical model, and the spin vector is
explicitly given by the function of time.  Making use of this solution
and applying a subtle time-dependent transformation of variables, the
system of equations are found to arrive at time-independent
homogeneous equations which can be exactly solved.  I also construct
all the possible POs.  With varying energy, those POs exhibit
peculiar interesting bifurcation scenarios, and all those
bifurcations are described analytically.
This kind of model would be not only valuable for the study of
nonlinear dynamics in integrable systems but also be suggestive for the
study of nonintegrable systems.

The paper is organized as follows.  Section~\ref{sec:spin_var}
recapitulates the basic concepts of the classical counterpart of the
spin and its dynamics derived by the spin coherent-state path-integral
representation.  In Sec.~\ref{sec:frozen_spin}, special types of
classical orbits with frozen spin and their properties are discussed.
In Sec.~\ref{sec:spin_prec}, generic solutions to the classical EOM
with spin precession are derived.  Using this solution, POs
are investigated and their bifurcation scenarios with varying
energies are clarified.  Section~\ref{sec:summary} is devoted to a
summary and discussion on the application of semiclassical trace
formula.

\section{Spin canonical variables and the classical equations of motion}
\label{sec:spin_var}

Spin is a quantum mechanical quantity analogous to the angular momentum.
In quantum mechanics, the spin operator $\hat{s}_i$ satisfies the following
angular momentum algebra,
\begin{equation}
[\hat{s}_i,\hat{s}_j]=i\hbar\eps_{i j k}\hat{s_k}, \label{eq:com_spin1}
\end{equation}
where $\eps_{i j k}$ stands for a completely antisymmetric unit tensor
(the Levi-Civita symbol), and the sum is taken over identical subscripts
in each term following the usual Einstein convention.
Spin raising and lowering operators
$\hat{s}_\pm\equiv \hat{s}_x\pm i\hat{s}_y$ satisfy
\begin{equation}
[\hat{s}_+,\hat{s}_-]=2\hbar\hat{s}_z, \quad
[\hat{s}_z,\hat{s}_\pm]=\pm\hbar\hat{s}_\pm. \label{eq:com_spin2}
\end{equation}
The classical counterpart of the spin is introduced via the
spin coherent state\cite{Kla79,KurSuz80} defined by
\begin{equation}
|\zeta\>=\frac{1}{(1+|\zeta|^2)^s}e^{\zeta\hat{s}_+/\hbar}|\sord,-\sord\>,
\end{equation}
where $\sord$ represents the spin quantum number\footnote{In the
following, I will use symbols with circle on top of them as the
dimensionless version of the variables without circle.  For example,
modulus of orbital angular momentum $l=\hbar\lord$ and spin-orbit
parameter $\kappa=\kbar\omega/s$, etc. in Sec.~\ref{sec:summary}.}
and $|\sord,m_s\>$ represents the spin eigenstate satisfying
\begin{gather}
\hat{s}^2|\sord,m_s\>=\hbar^2 \sord(\sord+1)|\sord,m_s\>, \nonumber \\
\hat{s}_z|\sord,m_s\>=\hbar m_s|\sord,m_s\>, \quad
m_s=-\sord,-\sord+1,\cdots,\sord.
\end{gather}
Here $\zeta$ is an arbitrary complex number and $|\zeta\>$ forms a
nonorthogonal, overcomplete set of spin state vectors.
The classical counterparts of the spin variables are defined
through the expectation values of the spin operators as
\begin{gather}
\begin{split}
s_x&=\<\zeta|\hat{s}_x|\zeta\>=s\sin\vartheta\cos\varphi, \\
s_y&=\<\zeta|\hat{s}_y|\zeta\>=s\sin\vartheta\sin\varphi, \\
s_z&=\<\zeta|\hat{s}_z|\zeta\>=s\cos\vartheta,
\end{split} \label{eq:spin_components}
\end{gather}
where the angles $\vartheta,\varphi$ are related to the complex
parameter $\zeta$ by
\begin{equation}
\frac{1}{\zeta}=\tan\frac{\vartheta}{2}e^{i\varphi}.
\end{equation}
The semiclassical approximation to the coherent state path integral
representation\cite{KurSuz80} derives the classical EOM as
\begin{gather}
\dot{\zeta}=\frac{(1+|\zeta|^2)^2}{2is}\pp{\sH}{\zeta^*}, \quad
\dot{\zeta}^*=-\frac{(1+|\zeta|^2)^2}{2is}\pp{\sH}{\zeta},
\end{gather}
with the Hamiltonian $\sH=\<\zeta|\hat{H}|\zeta\>$.
The EOM are translated to those for the spin variables as
\begin{equation}
\dot{s}_z=-\pp{\sH}{\varphi}, \quad \dot{\varphi}=\pp{\sH}{s_z},
\end{equation}
showing that the classical spin motion can be described by a
set of canonical variables $(q_s, p_s)=(\varphi, s_z)$.
Then, the Poisson brackets between spin components satisfy
\begin{equation}
\{s_i,s_j\}=\pp{s_i}{q_s}\pp{s_j}{p_s}-\pp{s_i}{p_s}\pp{s_j}{q_s}
=\eps_{i j k}s_k
\end{equation}
which is in exact correspondence with the quantum spin commutation
relation (\ref{eq:com_spin1}).

For the Hamiltonian (\ref{eq:Hamiltonian}), classical EOM are
expressed as
\begin{subequations}
\begin{gather}
\dot{\br}=\pp{H}{\bp}=\frac{\bp}{M}-\kappa \bs\times\br,
\label{eq:eom_r} \\
\dot{\bp}=-\pp{H}{\br}=-M\omega^2\br+\kappa \bp\times\bs,
\label{eq:eom_p} \\
\dot{\bs}=\{\bs,H\}=-\kappa(\br\times\bp)\times\bs.
\label{eq:eom_s}
\end{gather}
\end{subequations}
Time evolution of the orbital angular momentum $\bl=\br\times\bp$
becomes
\begin{gather}
\dot{\bl}=\kappa(\br\times\bp)\times\bs, \label{eq:eom_l}
\end{gather}
and the total angular momentum $\bj$ is shown to be
conserved:
\begin{equation}
\dot{\bl}+\dot{\bs}=0, \quad \bl+\bs=\bj=\mbox{const.}.
\end{equation}
Equations~(\ref{eq:eom_s}) and (\ref{eq:eom_l}) can be rewritten as
\begin{equation}
\dot{\bs}=-\kappa\bj\times\bs, \quad
\dot{\bl}=-\kappa\bs\times\bl=-\kappa\bj\times\bl,
\label{eq:spin_motion}
\end{equation}
and one sees that both $\bs$ and $\bl$ precess around the conserved
vector $\bj$ with angular velocity $\bm{\omega}_p=-\kappa\bj$ (see
Fig.~\ref{fig:precess}).

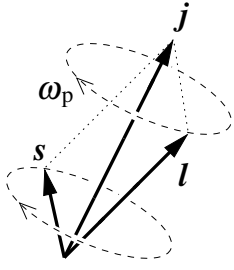
\putfig{tb}{.35}{precess}{Precession of angular momentum vectors.}

\section{Orbits with frozen spin}
\label{sec:frozen_spin}

By putting $\bj=(0,0,j)$, the spin components
satisfying Eq.~(\ref{eq:spin_motion}) are explicitly given by
the functions of time $t$ as
\begin{gather}
s_x=s_\perp\sin\omega_p t, \quad
s_y=s_\perp\cos\omega_p t, \label{eq:spin_precess} \\
(s_\perp=\sqrt{s^2-s_z^2}, \quad \omega_p=\kappa j), \nonumber
\end{gather}
with $s_z$ being constant.  First, let us consider the orbit with
$s_\perp=0$, for which spin vector $\bs=(0,0,s_z)$ is conserved.
In this case, the EOM become
\begin{gather}
\dot{x}=\frac{p_x}{M}+\kappa s_z y, \quad
\dot{p}_x=-M\omega^2 x+\kappa s_z p_y, \nonumber \\
\dot{y}=\frac{p_y}{M}-\kappa s_z x, \quad
\dot{p}_y=-M\omega^2 y-\kappa s_z p_x, \nonumber \\
\dot{z}=\frac{p_z}{M}, \quad
\dot{p}_z=-M\omega^2 z, \label{eq:eom_frozen}
\end{gather}
with $l_x=l_y=0$.
The motion in $z$ direction becomes a harmonic vibration with
angular frequency $\omega$.  Taking $x=y=0$, one obtains the
diametric PO
\begin{equation}
z(t)=z_0\sin\omega t, \quad z_0=\sqrt{\frac{2E}{M\omega^2}},
\end{equation}
with period $T=2\pi/\omega$.
Since the orientation of the spin (the choice of $z$ axis) is
arbitrary, the orbit forms a two-parameter family under
given energy $E$.

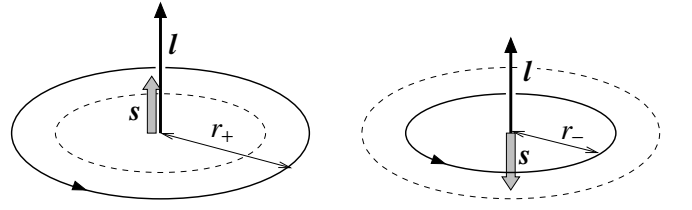
\putfig{tb}{}{circle}{Circular families of frozen-spin orbits.}

Alternatively, putting $z=0$ yields planar orbits in $x y$ plane.
To solve the EOM (\ref{eq:eom_frozen}) for $x$ and $y$ parts,
it is convenient to define the complex variables
\begin{equation}
\xi=x-\frac{i}{M\omega}p_x, \quad
\eta=y-\frac{i}{M\omega}p_y.
\end{equation}
The EOM (\ref{eq:eom_frozen}) are then transformed into
\begin{gather}
\mtrx{\dot{\xi}\\ \dot{\eta}}
=i\mtrx{\omega & -i\kappa s_z\\ i\kappa s_z & \omega}\mtrx{\xi \\ \eta}
\end{gather}
and one obtains two
eigenmodes with angular frequencies $\omega\pm\kappa s_z$.
For the frequency $\omega_c=\omega-\kappa s_z$, one has
\begin{gather}
\xi=i\eta=r_c e^{i\omega_c t}, \NN
x=\Re\xi=r_c \cos\omega_c t, \quad
y=\Re\eta=r_c \sin\omega_c t,
\label{eq:circ}
\end{gather}
which gives circular POs with angular frequencies
$\omega_c=\omega\mp \kappa s \equiv \omega_\pm$ corresponding to
$s_z=\pm s$.  From the EOM (\ref{eq:eom_frozen}), momentum components become
\begin{gather*}
p_x=M(\dot{x}-\kappa s_z y)=-M r_c\omega\sin\omega_c t, \\
p_y=M(\dot{y}+\kappa s_z x)=M r_c\omega\cos\omega_c t.
\end{gather*}
The orbital angular momentum $l_z$ and energy $E$ are then
expressed as
\begin{gather*}
l_z=x p_y-y p_x=M r_c^2\omega, \\
E=\frac{(M r_c\omega)^2}{2M}+\frac{M\omega^2 r_c^2}{2}-\kappa l_z s_z
 =M r_c^2\omega(\omega-\kappa s_z)=l_z\omega_c,
\end{gather*}
and the radii of the circular orbits are given by
\begin{equation}
r_c=\sqrt{\frac{l_z}{M\omega}}=\sqrt{\frac{E}{M\omega\omega_\pm}}
\equiv r_\pm.
\end{equation}
The angular frequencies $\omega_+$ and $\omega_-$ are of the
POs whose orbital angular
momenta are parallel and antiparallel to the spin,
respectively (see Fig.~\ref{fig:circle}).
Each of them forms a two-parameter family, as well as the diametric
family.

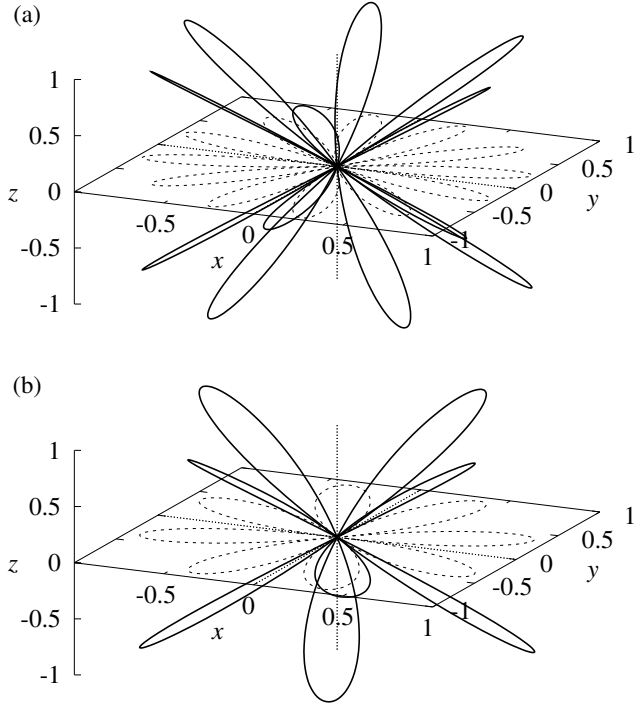
\putfig{tb}{}{frozen3d}{Examples of 3D frozen-spin orbits
for (a) $\kappa s=\omega/6$ and (b) $\kappa s=\omega/4$.}

In special cases where $\omega_+$ and $\omega_-$ are commensurable
with each other, all planar orbits become perisodic.  For a spin-orbit
parameter satisfying
\begin{equation}
\kappa s=\frac{m}{n}\omega,
\end{equation}
with integers $n$ and $m$ being relatively prime ($n>|m|\geq 1$),
the period of the orbit becomes
\begin{equation}
T=\frac{2\pi(n-m)}{\omega_+}=\frac{2\pi(n+m)}{\omega_-}
=\frac{2\pi n}{\omega}.
\end{equation}
In this case, $\omega_\pm$ is
also commensurable with $\omega$, and three-dimensional (3D) POs can
be realized by combining motion in the $z$ direction.  Such orbits
have $\bl=0$, for which the spin vector $\bs=\bj$ is conserved, and
the position vector $\br$ always pass through the origin
(see Fig.~\ref{fig:frozen3d}).

\section{Orbits with spin precession}
\label{sec:spin_prec}

\subsection{Bifurcations of circular orbits}

If the spin direction in each circular PO is slightly tilted,
the spin vector begins to precess around the
total angular momentum following the EOM (\ref{eq:spin_motion}).
Using Eq.~(24) and the EOM (\ref{eq:eom_frozen}), the
orbital angular momentum of the circular PO are obtained as
\begin{equation}
l_z=Mr_\pm^2\omega=\frac{E}{\omega_\pm}
\end{equation}
and the angular frequency of the spin precession is given by
\begin{equation}
\omega_p=\kappa j=\kappa\left(\frac{E}{\omega_\pm}\pm s\right).
\label{eq24}
\end{equation}
If this frequency becomes commensurable with that of the orbital
motion, $\omega_\pm$, a new PO will emerge in which orbital motion and
spin precession are combined.  The condition for such
bifurcation is given by
\begin{equation}
\omega_p=\frac{m}{k}\omega_\pm,
\label{eq25}
\end{equation}
where $m$ and $k$ denote mutually prime integers.
These conditions are successively satisfied by each circular PO
with varying energy $E$.  From Eqs.~(\ref{eq24}) and (\ref{eq25}),
the energies of these bifurcation points are given by
\begin{equation}
E^{\pm}_{k,m}=\frac{\omega_\pm(\frac{m}{k}\omega_\pm-\kappa s_z)}{\kappa},
\quad s_z=\pm s. \label{eq:bifpt_circle}
\end{equation}

\subsection{General solution to the equations of motion}

For a given total angular momentum $\bj=(0,0,j)$, the EOM for the spin
(\ref{eq:spin_motion}) can be solved, and the spin components are
explicitly given as functions of time by Eqs.~(\ref{eq:spin_precess}).
Using this solution, the EOM for the orbital degrees of freedom
are expressed as
\begin{gather}
\begin{array}{rl}
\dot{x}&=\dfrac{p_x}{M}-\kappa(z s_y-y s_z), \\[6pt]
\dot{y}&=\dfrac{p_y}{M}+\kappa(z s_x-x s_z), \\[6pt]
\dot{z}&=\dfrac{p_z}{M}+\kappa(x s_y-y s_x), \\[6pt]
\dot{p}_x&=-M\omega^2 x-\kappa(p_z s_y-p_y s_z), \\
\dot{p}_y&=-M\omega^2 y+\kappa(p_z s_x-p_x s_z), \\
\dot{p}_z&=-M\omega^2 z+\kappa(p_x s_y-p_y s_x)
\end{array}
\label{eq:eom_general}
\end{gather}
The task is to find a solution to the above coupled linear equations
for $\br$ and $\bp$ with time-dependent coefficients $s_x(t)$ and
$s_y(t)$ given by Eq.~(\ref{eq:spin_precess}).  By introducing complex
variables
\begin{equation}
\xi=x-\frac{i p_x}{M\omega}, \quad
\eta=y-\frac{i p_y}{M\omega}, \quad
\zeta=z-\frac{i p_z}{M\omega},
\end{equation}
the EOM (\ref{eq:eom_general}) are transformed into a compact form as
\begin{gather}
\begin{array}{rl}
\dot{\xi}&=i\omega\xi-\kappa(\zeta s_y-\eta s_z), \\
\dot{\eta}&=i\omega\eta+\kappa(\zeta s_x-\xi s_z), \\
\dot{\zeta}&=i\omega\zeta+\kappa(\xi s_y-\eta s_x).
\end{array}
\label{eq:eom_gen2}
\end{gather}
The energy $E$ and component $l_z$ of the orbital angular momentum
are expressed as
\begin{gather}
E=\frac{M\omega^2}{2}(|\xi|^2+|\eta|^2+|\zeta|^2)
-\kappa(l_zs_z-s_\perp^2), \\
l_z=x p_y-y p_x=\frac{i M\omega}{2}(\xi^*\eta-\xi\eta^*).
\end{gather}
Because of the relation $\bj=\bl+\bs=(0, 0, j)$, the $x$ and $y$ components
of the orbital angular momentum satisfy $l_x=-s_x$ and $l_y=-s_y$.
By a further transformation $(u,v)=(\xi+i\eta,\xi-i\eta)$ which replaces
the trigonometric functions in the coefficients of EOM with exponential
functions, one obtains the relations
\begin{gather}
E=\frac{M\omega^2}{2}\left(\frac{|u|^2+|v|^2}{2}+|\zeta|^2\right)
 -\kappa(l_zs_z-s_\perp^2), \label{eq:E_uv} \\
l_z=\frac{M\omega}{4}(|u|^2-|v|^2), \label{eq:lz_uv}
\end{gather}
and the EOM (\ref{eq:eom_gen2}) appear as
\begin{gather}
\begin{array}{rl}
\dot{u}&=i(\omega-\kappa s_z) u-\kappa s_\perp\zeta e^{-i\omega_p t}, \\
\dot{v}&=i(\omega+\kappa s_z) v-\kappa s_\perp\zeta e^{i\omega_p t}, \\
\dot{\zeta}&=i\omega\zeta+\tfrac12\kappa
  s_\perp(u e^{i\omega_p t}+v e^{-i\omega_p t}).
\end{array}
\label{eq:eom_gen3}
\end{gather}
Putting $u e^{i\omega_p t}=U$ and $v e^{-i\omega_p t}=V$, one finally
obtains coupled EOM with constant coefficients as
\begin{gather}
\begin{array}{rl}
\dot{\zeta}&=i\omega\zeta+\tfrac12\kappa s_\perp(U+V), \\
\dot{U}-i\omega_p U&=i(\omega-\kappa s_z)U-\kappa s_\perp\zeta, \\
\dot{V}+i\omega_p V&=i(\omega+\kappa s_z)V-\kappa s_\perp\zeta,
\end{array}
\label{eq:eom_gen4a}
\end{gather}
that are combined into a normal mode equation
\begin{align}
&\frac{d}{dt}\mtrx{\zeta\\ U\\ V}
=i\bM\mtrx{\zeta\\ U\\ V}, \nonumber \\
&\bM=\mtrx{\omega & -i\kappa s_\perp/2 & -i\kappa s_\perp/2 \\
i\kappa s_\perp & \omega+\omega_p-\kappa s_z & 0  \\
i\kappa s_\perp & 0 & \omega-\omega_p+\kappa s_z}.
\label{eq:eom_gen4}
\end{align}
By solving the secular equation $\det(\mu-\bM)=0$, the three
eigenvalues $\mu$ of the matrix $\bM$ are obtained as
\begin{equation}
\mu=\omega, \quad \omega\pm\kappa l, \label{eq:mu_eigen}
\end{equation}
where I use the relation
\begin{gather*}
(\omega_p-\kappa s_z)^2+(\kappa s_\perp)^2
=(\kappa l_z)^2+(-\kappa l_\perp)^2=(\kappa l)^2.
\end{gather*}
The generic orbit can then be expressed by a linear combination of
these three modes:
\begin{equation}
\mtrx{\zeta\\ U\\ V}=\left\{
\mtrx{Z_0\\ U_0\\ V_0}
+\mtrx{Z_+\\ U_+\\ V_+}e^{i\kappa l t}
+\mtrx{Z_-\\ U_-\\ V_-}e^{-i\kappa l t}
\right\}e^{i\omega t}, \label{eq38}
\end{equation}
with the coefficients satisfying the following relation:
\begin{gather}
U_0=-V_0=\frac{is_\perp}{l_z}Z_0, \NN
U_{\pm}=\frac{is_\perp}{\pm l-l_z}Z_\pm=-\frac{\pm l+l_z}{is_\perp}Z_\pm,
\NN
V_{\pm}=\frac{is_\perp}{\pm l+l_z}Z_\pm=-\frac{\pm l-l_z}{is_\perp}Z_\pm.
\label{eq39}
\end{gather}
In the above, $l_z\ne 0$ is assumed, which is satisfied by most of the
orbits under consideration.  (Possible POs with $l_z=0$
are described in the Appendix.)
By applying Eqs.~(\ref{eq38}) and (\ref{eq39}) to Eq.~(\ref{eq:lz_uv}),
one obtains the relation
\begin{align}
l_z&=\frac14M\omega\left(|U|^2-|V|^2\right) \NN
&=M\omega\left[\frac{l\,l_z}{s_\perp^2}(|Z_+|^2-|Z_-|^2) \right. \NN
&\hspace{8ex} \left.
 +\frac{l}{l_z}\Re\Bigl\{(Z_0^*Z_+-Z_0Z_-^*)
 e^{i\kappa l t}\Bigr\}\right].
\end{align}
In order that $l_z$ be independent of time, one must have
\[
Z_0^*Z_+-Z_0Z_-^*=0.
\]
Nonvanishing $Z_0$ leads to $|Z_+|=|Z_-|$ and thus to the
improper result $l_z=0$.  Therefore, $Z_0=0$ and one obtains
\begin{gather}
|Z_+|^2-|Z_-|^2=\frac{s_\perp^2}{M\omega l}. \label{eq:zz_relation1}
\end{gather}

\subsection{Bridge between circular orbits}

The periodicity condition is now considered.
There are three independent angular frequencies;
$\omega_p$ for the spin precession, and $\omega\pm\kappa l$
for the orbital motion in the $z$ direction.  The frequencies for the motion
in $x$ and $y$ directions ($\omega\pm\kappa l\pm\omega_p$) are
given by the combinations of the above three.
Let us first consider the case where one of the coefficients
$Z_\pm$ vanishes.
In such a case, only one of the two frequencies $\omega\pm\kappa l$
should be considered in the periodicity condition.
Taking into account the relation (\ref{eq:zz_relation1}),
let us put
\begin{equation}
Z_-=0, \quad Z_+=\frac{s_\perp}{\sqrt{M\omega l}}e^{i\gamma},
\label{eq42}
\end{equation}
where $\gamma$ is an arbitrary phase parameter.
The periodicity condition then requires
\begin{equation}
\omega+\kappa l=\frac{n}{m}\omega_p,
\label{eq:bridge_mu}
\end{equation}
where $n$ and $m$ are relatively prime integers satisfying $n>m$, and the 
period of the PO is given by
\begin{equation}
T=\frac{2\pi m}{\omega_p}=\frac{2\pi n}{\omega+\kappa l}
 =\frac{2\pi(n\pm m)}{\omega+\kappa l\pm\omega_p}.
\label{eq:bridge_T}
\end{equation}
Rewriting Eq.~(\ref{eq:bridge_mu}) in terms of $j$ and $s_z$ yields
\begin{equation}
\omega+\kappa\sqrt{j^2-2js_z+s^2}=\frac{n}{m}\kappa j,
\end{equation}
with which $j$ can be expressed as a function of $s_z$ for each set
of $(n,m)$,
\begin{align}
\kappa j=&\frac{m}{n^2-m^2}\Bigl(n\omega-m\kappa s_z \NN
&+\sqrt{(n\omega-m\kappa s_z)^2-(n^2-m^2)(\omega^2-\kappa^2 s^2)}
\Bigr).
\end{align}
Then, one can determine $\omega_p=\kappa j$, $l_z=j-s_z$,
$l=\sqrt{l_z^2+s_\perp^2}$, and the coefficients $U_+$, $V_+$
via Eq.~(\ref{eq42}) that are necessary to
calculate the PO for each value of $s_z$.
Using these quantities, the energy (\ref{eq:E_uv}) of the PO is
obtained as
\begin{equation}
E=\omega\sqrt{j^2-2js_z+s^2}-\kappa(js_z-s^2). \label{eq:e_circ}
\end{equation}

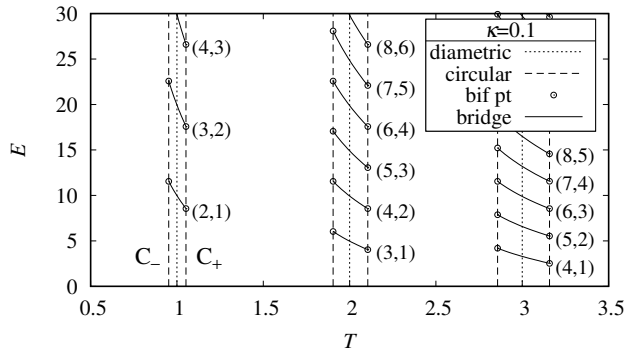
\putfig{tb}{}{etplot}{Plot of the energy $E$ (in units of $\hbar\omega$)
vs. period $T$ (in units of $2\pi/\omega$) of the POs.
The vertical lines represent the diametric orbits (dotted line) and
two types of circular orbits $C_\pm$ (dashed lines).
Open circles indicate the bifurcation points
(\ref{eq:bifpt_circle}), and the curves connected them
represent the bridge orbits $(n,m)$.}

Figure~\ref{fig:etplot} shows the relation between energy $E$ and
period $T$ of the above PO for various sets of $(n,m)$ with varying
$s_z$ from $s$ to $-s$.  Each such PO
forms a bridge between the two circular orbits, as it emerge from
the circular orbit C$_+$, increasing spin obliquity with energy and
finally submerge into another circular orbit C$_-$.
It can be easily checked that the above orbit coincides the
circular orbits (\ref{eq:circ}) in the limits $s_z\to \pm s$.
The bifurcation scenario of the bridge orbit is illustrated in
Fig.~\ref{fig:brhis}.

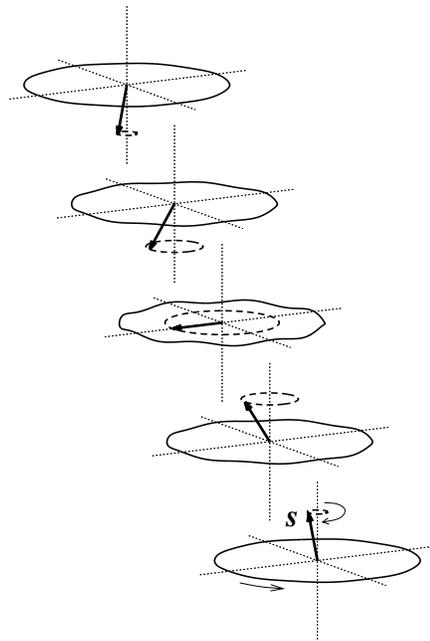
\putfig{tb}{.7}{brhis}{Illustration of the bifurcation scenario of
the bridge orbit.  Lower to upper panels in energy order.}

Each bridge orbit forms a three-parameter family
since the orientation of the total angular momentum $\bj$ and the phase
$\gamma$ in Eq.~(\ref{eq42}) are arbitrary.
Figure~\ref{fig:bridge} displays some examples of
bridge orbits $(n,m)$ for several values of $s_z$.  As the spin vector
tilts, the orbital ripples develop and eventually subside toward
the spin flip.

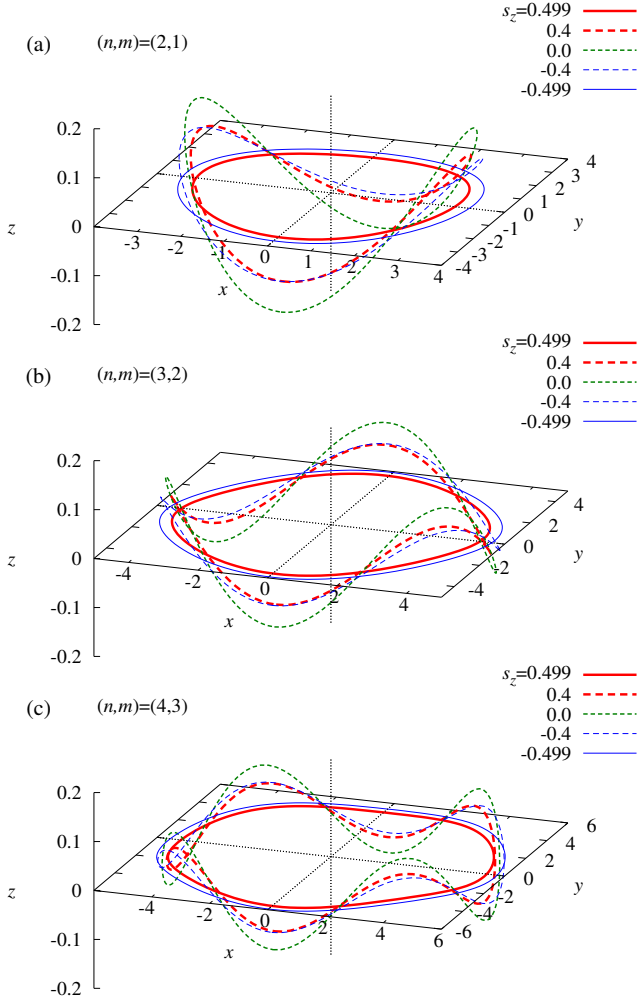
\putfig{tb}{}{bridge}{Some bridge orbits $(n,m)$ with
$n-m=1$, calculated for
several values of $s_z$ (in units of $\hbar$) between $s$ and $-s$.
The parameters $\kappa=0.1\,\omega/\hbar$ and $s=\hbar/2$ are used.
Note that the vertical magnification is set approximately ten times the
horizontal one.}

\subsection{Secondary bifurcation}

The period of each bridge orbit $(n,m)$ always coincides with a multiple
of $2\pi/\omega$ somewhere on the way from $C_+$ to $C_-$ with
increasing energy.
Since $\omega+\kappa l$ becomes commensurable with $\omega$ at that
point, it is also commensurable with $\omega-\kappa l$ and the
combination of the modes $\mu=\omega\pm\kappa l$
\begin{equation}
\mtrx{\zeta \\ u e^{i\kappa j t}\\ v e^{-i\kappa j t}}
=\left\{\mtrx{Z_+ \\ U_+ \\ V_+}e^{i\kappa l t}
 +\mtrx{Z_- \\ U_- \\ V_-}e^{-i\kappa l t}\right\}e^{i\omega t}
\label{eq48}
\end{equation}
provides the periodic solution.
The commensurability condition
\begin{equation}
\frac{\omega+\kappa l}{n}=\frac{\kappa j}{m}=\frac{\omega}{k}
\end{equation}
and trigonometric inequality $|j-l|\leq s$ lead to the following
relation:
\begin{equation}
|j-l|=\left|j-\frac{n-k}{m}j\right|=\frac{|m-(n-k)|}{m}j\leq s.
\label{eq50}
\end{equation}
This condition is satisfied at least once for all $(n,m)$
at $k=n-m$.
For sufficiently small $j$ and large $m$ satisfying $ms/j > 1$,
one finds several integers $k$ which satisfy (\ref{eq50}),
but solutions other than $k=n-m$ correspond to rather long
orbits with period $T>2\pi/\kappa s$ and are irrelevant to the
gross shell structure.

The PO is given by Eq.~(\ref{eq48}) with coefficients
satisfying Eqs.~(\ref{eq39}) and (\ref{eq:zz_relation1}).
By inserting
these relations into Eq.~(\ref{eq:E_uv}), the energy is expressed as
\begin{gather}
E=\frac{M\omega^2l^2}{s_\perp^2}(|Z_+|^2+|Z_-|^2)+\frac12\kappa s^2.
\label{eq:zz_relation2}
\end{gather}
The orbit bifurcated from the bridge $(2,1)$ is displayed
in Fig.~\ref{fig:bbif} for several values of energy.

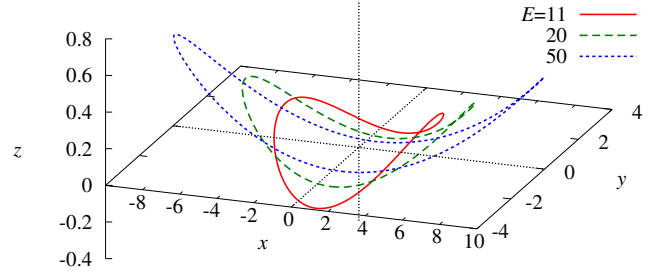
\putfig{tb}{}{bbif}{PO bifurcated from the
bridge $(n,m)=(2,1)$ at
several values of energy $E$ (in units of $\hbar\omega$).
$\kappa=0.1\,\omega/\hbar$, $s=\hbar/2$ are taken,
and $E_{\rm bif}=10.025\,\hbar\omega$.}

Using Eqs.~(\ref{eq:zz_relation1}) and
(\ref{eq:zz_relation2}), one obtains
\begin{gather}
|Z_\pm|^2=\frac{s_\perp^2}{2M\omega^2l^2}\left\{E-\tfrac12\kappa s^2
 \pm\omega l\right\}.
\end{gather}
The bifurcation energy corresponding to $|Z_-|=0$ is given by
\begin{equation}
E_{\rm bif}^{(n,m,k)}=\frac{n-k}{k}\frac{\omega^2}{\kappa}
+\frac12\kappa s^2,
\end{equation} 
and the orbit exists for $E\geq E_{\rm bif}$.
Each orbit $(n,m,k)$ forms a four-parameter family
since the orientation of $\bj$ and the arguments $\gamma_\pm$ of the
complex coefficients $Z_\pm=|Z_\pm|e^{i\gamma_\pm}$ are arbitrary.
As displayed in the $E$-$T$ plots in Fig.~\ref{fig:etplot2},
POs with the same period $T=2\pi k/\omega$ but having different
geometries $(n,m,k)$ successively emerge with increasing energy.

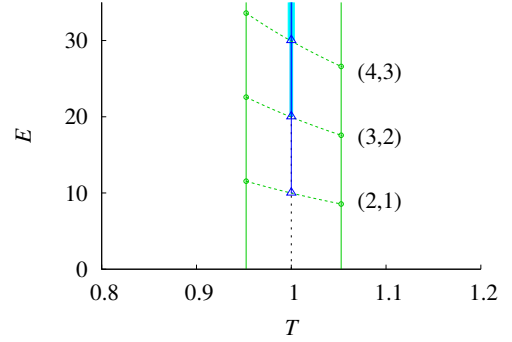
\putfig{tb}{.8}{etplot2}{$E$-$T$ plot for the orbits bifurcated
from the middle of the bridge orbits.  The solid and broken lines
represent the circular orbits and bridge between them, respectively.
The open triangles indicate the bifurcation points of the bridges, and
the straight lines on the top of them represent the new orbits.
The units for $T$ and $E$ are same as in Fig.~\ref{fig:etplot}.}

\section{Summary and discussion}
\label{sec:summary}

Classical motion was investigated in the extended phase space for the
system of a spherical harmonic oscillator with simple spin-orbit
coupling.  General solutions to the EOM and all the POs
are obtained analytically, and the bifurcation scenarios are fully
clarified in both generic and specific values of the spin-orbit
parameter.  For arbitrary value of the spin-orbit parameter, there are
two kinds of degenerate two-parameter families of POs, diametric and
circular ones, with frozen spin.
There are two circular families C$_\pm$ in which spin is
parallel or antiparallel to the orbital angular momentum.  These
circular orbits encounter successive bifurcations with increasing
energy and generate three-parameter bridge PO families.  On its way
from one circular PO to another, each bridge encounters another
bifurcation and the four-parameter PO family emerges, which has a
constant period and survives for any higher energy.  For specific
values of spin-orbit parameter, different types of periodic solutions
are also possible, for both frozen and moving spins.

Classical-quantum correspondence for this system was considered in
Ref.~\cite{AmaBra02}.
In the direction from quantum to classical,
the authors constructed the exact trace formula analogous to the
Berry-Tabor formula for multiply periodic systems\cite{BerTab76}.
Quantum energy eigenvalues are given explicitly in terms of
the quantum numbers as
\begin{gather}
E_{n\lord\jord}=\left\{
\begin{array}{l@{\quad}l}
 \hbar\omega(2n+\lord+\tfrac32-\kbar \lord) & (\jord=\lord+1/2) \\
 \hbar\omega(2n+\lord+\tfrac32+\kbar(\lord+1)) & (\jord=\lord-1/2)
\end{array}\right.,
\label{eq:spectrum} \\
(\kbar=\kappa s/\omega, \quad
n=0,1,2,\cdots, \quad \lord=0,1,2,\cdots) \nonumber
\end{gather}
where $n, \lord, \jord$ are radial, orbital angular momentum,
and total angular momentum quantum numbers, respectively.
Then, the level density is expressed as
\begin{equation}
g(E)=\sum_{n,\lord,\jord=\lord\pm 1/2}(2\jord+1)\delta(E-E_{n\lord\jord}).
\end{equation}
By applying the Poisson's summation formula to the above double sum,
the level density is transformed into the form of summation over
PO contributions as
\begin{equation}
g(E)=\bar{g}(E)+\sum_{\rm po}A_{\rm po}\cos\left(\tfrac{1}{\hbar}
T_{\rm po}E-\tfrac{\pi}{2}\mu_{\rm po}\right).
\label{eq:trace_BT}
\end{equation}
The obtained formula includes contributions of POs with three
primitive periods,
\[
T_0=\frac{2\pi}{\omega}, \quad T_\pm=\frac{2\pi}{\omega\mp\kappa s}.
\]

In the direction from classical to quantum,
application of the WKB method was considered in \cite{AmaBra02}.
By diagonalizing the Hamiltonians with
respect to the spin channels, two adiabatic Hamiltonians $H_\pm=H_0\pm
\kappa s|\bm{l}|$ are obtained.  Circular
orbits with periods $T_\pm$ are found in the Hamiltonian $H_\pm$,
respectively, which are equivalent to the circular family with frozen
spin discussed in Sec.~\ref{sec:frozen_spin}.  Straight-line orbits
with period $T_0$ are also found in both $H_\pm$ but they have
$\bm{l}=0$ and lie on the mode conversion point $H_+=H_-$, hence the
contribution to the level density cannot be evaluated.  Apart from the
mode conversion problem, they noticed another difficulty in
classical-quantum correspondence.  Degeneracy of the PO family is
related to the semiclassical order of its contribution to the level
density.  In general, the amplitude factor $A_{\rm po}$ in the trace
formula is proportional to $\hbar^{-K/2}$ for the $K$-parameter
families.  In the exact trace formula, the lowest order term in the
amplitude $A_{\rm po}$ for PO with periods $T_\pm$ is proportional to
$\hbar^{-2}$ suggesting $K=4$ families.  However, the circular
orbits in $H_\pm$ form only $K=2$ families and the
semiclassical order of their contributions are inconsistent.

Then, how about the Hamiltonian in the extended phase space taking the
spin motion into account.  The authors of Ref.~\cite{AmaBra02} seem
to have noticed about the existence of frozen-spin circular and diametric
POs with period $T_\pm$ and $T_0$, respectively, but they didn't
discuss the matter further, probably because they do not have
enough information on the classical POs.
Now we have information on all the classical POs in the
extended phase space.  Unfortunately, the
problem in semiclassical order of the contributing POs are not
simply solved with the present result. 
The orbits having the periods $T_\pm$ are only the frozen-spin circular
orbits that form $K=2$ families as well as in the WKB method.
However, they encounter successive bifurcations with varying
energy, which might enhance the contribution to the level density.
Bridge orbits between circular POs have $K=3$ and their
periods vary from $T_+$ and $T_-$ with increasing energy, but their
contributions are not existing in the current form of the exact
trace formula (\ref{eq:trace_BT}).
There are two kinds of POs having the period $T_0$: $K=2$ family of
diametric orbit with frozen spin, and $K=4$ family of `4D' orbits
which emerge from the second bifurcations of the bridge POs.
In the exact formula, the amplitude of PO with period $T_0$ suggests
a $K=2$ family, hence a kind of cancellation between the $K=4$ families
seems to be necessary.

It is an interesting open question if the above difficulties can be
overcome and a more appropriate quantum-classical correspondence can
be obtained by the use of trace formula in the extended phase space
with suitable uniform approximations\cite{KaiBra04b,AriBra08a}, and
also by another way of resummation of quantum numbers from the quantum
spectrum (\ref{eq:spectrum}) to the formula (\ref{eq:trace_BT}).

\appendix
\def\thesection{\relax}
\section{Other minor orbits with spin precession}

For the sake of completeness, I consider here the $l_z=0$ POs
with spin precession, which are not considered in the main part
because of their insignificant contribution to the trace formula.
Note that $j=s_z$ and $l=s_\perp$,
and the coefficients in Eq.~(\ref{eq38}) are related as
\begin{gather*}
Z_0=0, \quad U_0=-V_0, \quad
U_\pm=V_\pm=\pm i Z_\pm.
\end{gather*}
Inserting these into the conditions
\begin{gather*}
l_z=\frac{M\omega}{4}(|u|^2-|v|^2)=0, \\
l_x+i l_y=\frac{M\omega}{2}(\zeta v^*-\zeta^* u)
=-i s_\perp e^{-i\omega_p t}, \\
E=\frac{M\omega^2}{2}\left(\frac{|u|^2+|v|^2}{2}+|\zeta|^2\right)
 +\kappa s_\perp^2,
\end{gather*}
one has
\begin{gather}
U_0=0, \quad |Z_+|^2-|Z_-|^2=\frac{s_\perp}{M\omega}, \NN
|Z_+|^2+|Z_-|^2=\frac{E-\kappa s_\perp^2}{M\omega^2}.
\label{eq:lzero_coeff}
\end{gather}
For $Z_-=0$, the periodicity condition is
\begin{gather*}
\frac{\kappa s_z}{m}=\frac{\omega+\kappa s_\perp}{n}
\end{gather*}
and $(s_z, s_\perp)$ are determined as function of $\kappa$ for given
set of $(n, m)$.  The period is given by
\[
T=\frac{2\pi n}{\omega+\kappa s_\perp}=\frac{2\pi m}{\kappa s_z},
\]
where $n>\omega/\kappa s$ becomes a large integer for a typical value
of $\kappa$ for atomic nuclei ($\kbar\approx 0.06$) and corresponds to a
long PO.  As shown from Eq.~(\ref{eq:lzero_coeff}), this orbit can
exist only at the energy
\[
E=\omega s_\perp+\kappa s_\perp^2
\]
and cannot contribute to the shell structure.
With varying spin-orbit parameter $\kappa$, it disappears in the
limit $s_\perp\to 0$ ($\kappa\to m\omega/ns$).

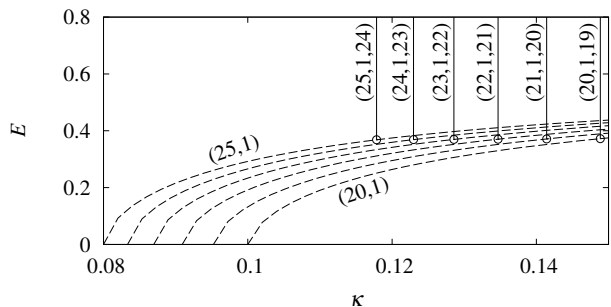
\putfig{tb}{}{lzero}{$l_z=0$ PO with spin precession.
Broken curves represent $Z_-~0$ orbit $(n,m)$ with $m=1$, and the
solid lines on top of the bifurcation point marked by open circles
represent $Z_-\ne 0$ orbit $(n,m,k)$.
Energy of PO (in units of $\hbar\omega$)
is plotted as a function of the LS coupling strength $\kappa$ (in units
of $\omega/\hbar$).
Broken curves represent the PO with $Z_-=0$, and the open dots represent
bifurcation points.  At those values of spin-orbit parameter $\kappa$,
the 4 parameter PO families with $Z_-\ne 0$ having constant
periods $T=2\pi k/\omega$ ($k<n$) emerge, and they survive for higher
energies.}

For $Z_-\ne 0$, the periodicity condition becomes
\begin{gather*}
\frac{\kappa s_z}{m}=\frac{\omega+\kappa s_\perp}{n}
=\frac{\omega-\kappa s_\perp}{n'}=\frac{\omega}{k} \quad (n'=2k-n), \\
\quad \frac{s_z}{s_\perp}=\frac{m}{n-k},
\end{gather*}
and the period is given by
\[
T=\frac{2\pi k}{\omega}.
\]
This condition is satisfied only by specific values of $\kappa$ and
certain spin orientations for which $\kappa s_z$ and $\kappa s_\perp$
are both commensurable with $\omega$, but once the orbit emerges, it can
exist for any higher energy $E$ with the coefficients
\begin{gather*}
|Z_\pm|^2=\frac12\left(\frac{E-\kappa s_\perp^2}{M\omega^2}
 \pm\frac{s_\perp}{M\omega}\right), \\
E\geq E_{\rm bif}=\omega s_\perp+\kappa s_\perp^2.
\end{gather*}
$k=m\omega/\kappa s_z>\omega/\kappa s$ is a large integer, which
again gives a long PO.  Figure~\ref{fig:lzero} shows where the shortest
$l_z=0$ POs are found that exist around $\kappa=0.12\,\omega/\hbar$.  In the
case of $(n,m,k)=(25,1,24)$, for instance, one sees
\begin{gather*}
s_z=s_\perp=\frac{s}{\sqrt{2}}, \\
\kappa=\frac{\omega}{24s_z}\fallingdotseq 0.118\,\frac{\omega}{\hbar}, \quad
E_{\rm bif}\fallingdotseq 0.368\,\hbar\omega.
\end{gather*}

\bibliographystyle{apsrev4-2}
\bibliography{refs_spin}
\end{document}